\documentclass[showpacs,preprintnumbers,amsmath,amssymb]{revtex4}
\usepackage{graphicx}

\begin{document}
\title{Inverse melting and  inverse freezing: a spin model}
\author{Nurith Schupper and Nadav M. Shnerb}
\affiliation{Department of Physics, Bar-Ilan University, Ramat-Gan
52900 Israel}

\begin{abstract}
Systems of highly degenerate ordered or frozen state may exhibit
inverse melting (reversible crystallization upon heating) or
inverse freezing (reversible glass transition upon heating). This
phenomena is reviewed, and a list of experimental demonstrations
and theoretical models is presented. A simple spin model for
inverse melting is introduced and solved analytically for infinite
range, constant paramagnetic exchange interaction. The  random
exchange analogue of this model yields inverse freezing, as
implied by the analytic solution based on the replica trick. The
qualitative features of this system (generalized Blume-Capel spin
model) are shown to resemble a large class of inverse melting
phenomena. The appearance of inverse melting is related to an
exact rescaling of one of the interaction parameters that measures
the entropy of the system. For the case of almost degenerate spin
states perturbative expansion is presented, and the first three
terms correspond to the empiric formula for the Flory-Huggins
$\chi$ parameter in the theory of polymer melts. Possible
microscopic origin of this $\chi$ parameter and the limitations of
the Flory-Huggins theory where the state degeneracy is  associated
with the different conformations of a single polymer or with the
spatial structures of two interacting molecules are discussed.
\end{abstract}


\maketitle

\section{introduction}

Inverse melting is a reversible transition between a liquid phase
at low temperatures to a high temperature crystalline phase. This
is an unusual and counter-intuitive phenomenon in which isobaric
addition of heat causes liquids to crystallize, the reverse of the
usual situation. For the opposite process, of melting a solid to a
liquid, which is normally expected to produce cooling, in inverse
melting heat is released as the solid melts. Clearly, inverse
melting happens if, and only if, the so called "ordered" phase
(crystal) admits more entropy than the "disordered" state; this
may occur, e.g., if in the liquid phase some of the degrees of
freedom of the elementary constituents are frozen, and melt in the
crystalline phase.

Speaking about freezing and crystalization one should make the
distinction between static phenomena, such as magnetization,
degree of phase separation and crystalline order (Bragg peaks),
and dynamical aspects, like the response functions, viscosity,
ergodicity breakdown  and so on. While the static features reflect
the properties of the "ground state" (lowest free energy state),
the dynamics is dictated by the size of the potential barriers
among different states. In the following, cases where the
appearance of crystalline order is correlated with higher response
functions will be mentioned, along with situations where
ergodicity breaks down without the appearance of ordered
structure, i.e., glass like transition. If such a transition
occurs upon temperature increase we are speaking about "inverse"
glass transition, or inverse freezing, analogous to inverse
melting.

Although rare, real substance examples of inverse melting
phenomena have been found in a wide range of systems, as well as
the formation, upon heating, of solid amorphous or glassy states.
Since the transition that occurs on heating absorbs heat (as does
normal melting), and the phase in equilibrium at higher
temperatures has higher disorder or entropy,  the crystalline or
frozen amorphous phase are more disordered than the liquid phase.
In each case, the greater order of average atomic positions in the
crystal has to be offset by greater disorder in some other
characteristic. Thus, all cases of inverse melting and inverse
glass transition appear to involve a freezing of the center of
mass location of the system constituents (molecules, polymers,
flux lines). The loss of entropy due to this freezing is
compensated by other  microscopic degrees of freedom that are
coupled to the center-of-mass position, where localization of the
center of mass  increases the amount of such excitations.

The aim of this paper is to survey the literature concerning the
subject, to present and discuss an extremely simple spin model for
inverse melting and inverse freezing, a model that has been
recently presented by the authors \cite{schupper},  and to extract
some general features related to the phenomenon. The paper is
organized as follows: in the second section we discuss the general
characteristics of inverse melting and inverse glass transition
and classify the different possible inverse melting scenarios. In
section 3 we survey some real material examples from the
literature, and ascribe them to the different classifications
presented. Section 4 is devoted to modelling, where our simple
spin model based on the well known Blume-Capel \cite{Blume} model
is presented along with previously discussed models. In section 5
we deal with the ordered version of our "enriched" Blume-Capel
model for inverse melting, and in the next section its disordered
version and inverse freezing are discussed. Section 7 deals with
the scaling properties of the model and we also  show how its
modification can be related to the temperature dependence of the
interaction parameter of the Flory-Huggins like  \cite{flory}
\cite{Huggins} theories. Finally, some general conclusions and
remarks are presented.

\section{Inverse melting and inverse freezing scenarios}

\subsection{Two types of first order inverse melting}

Following Stillinger and Debenedetti, \cite{stillinger}, we begin
the discussion of inverse melting with the Clausius-Clapeyron
equation that describes the slope of the melting curve in first
order transition, e.g.,  the curve that describes the boundary
between a crystal and a liquid in the T-P plane

\begin{eqnarray}
\label{eq:Clausius} \frac{d
P}{dT}=\frac{S^{(2)}-S^{(1)}}{V^{(2)}-V^{(1)}}
\end{eqnarray}

\begin{figure}
  \includegraphics[width=7.7cm]{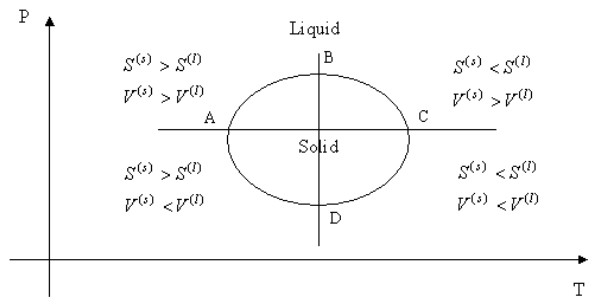}\\
  \caption
{A sketch of the different melting curves in the T-P plane
resulting from the Clausius-Clapeyron equation, including both
types of inverse melting scenarios. } \label{fig1}
\end{figure}

where $P(T)$ is the temperature dependent melting pressure, $S$
and $V$ denote the molar entropies and volumes, and the
superscripts $2$ and $1$ denote the high and low temperature
phases, respectively. Alternatively, these will be denoted by $l$
and $s$ for the liquid and solid phase. As noted first by Tammann
\cite{Tammann}, this thermodynamic equation offers schematically 4
different types of melting curves as shown in Figure (\ref{fig1})
and will help to classify the known examples of \emph{first order
transition} scenarios. If the liquid-crystal  transition is first
order, at least one of $S^{(l)}-S^{(s)}$ and $V^{(l)}-V^{(s)}$ is
non-zero at every point of the curve.

Let us identify the different regimes in this diagram. 'Normal'
melting involves an increase in both the entropy (the system
absorbs latent heat) and the molar volume as the crystal becomes
liquid. In that case, both $S^{(2)}-S^{(1)}$ and $V^{(2)}-V^{(1)}$
are positive, and therefore the slope of the curve is also
positive. In figure (\ref{fig1}) this is the portion of the  curve
between the points C and D. "Anomalous", or water-like, melting
happens if the molar volume of the liquid is larger than that of
the solid, $V^{(2)}-V^{(1)}$ becomes negative and the slope of the
first order transition curve is also negative, i.e., the melting
temperature decreases as pressure increases. The curve between the
points B and C demonstrate this situation.

The left side of the circle, i.e., the  intervals between  A and B
and between  A and D are ranges of inverse melting, where isobaric
heating takes the system from its liquid phase into the
crystalline phase. If the transition involves latent heat, for any
inverse melting situation $S^{(s)}-S^{(l)}>0$.   For the interval
from A to D  $V^{(s)}-V^{(l)}$ is negative (the solid volume is
smaller than the liquid) while the interval AB exhibits positive
slope, since the solid is less dense than the liquid. Thus, there
are two types of inverse melting, similar to the normal and
anomalous usual melting. We shall denote the former case as
inverse melting of type I, while the "anomalous" case of larger
molar volume in the ordered phase will be denoted as inverse
melting of type II.

\subsection{Order-disorder transition and response functions}

As discussed in the introduction, the "standard" solid-liquid
transition involves both static (symmetry breakdown, Bragg peaks)
and dynamic (diverging viscosity, rapid changes in the Young
modulus, discontinuous susceptibility) aspects. In general, any of
these may take place independently.  For example, one may find an
amorphous system with diverging (or at least very large)
viscosity, like a glass. On the other hand, the order parameter
may take a finite value but the ordered system is "softer" than
the disordered one, i.e., its response functions are larger, and
therefore its viscosity smaller. In general we will speak about
inverse melting when a liquid acquires crystalline structure upon
heating, and about inverse freezing if the liquid becomes a glass
(amorphus solid, with higher viscosity or an increase in other
response functions, but no apparent order). An inverse
solid-amorphous transition is another situation where an amorphous
rigid material (similar to window glass) reversibly crystalizes as
its temperature increases.

Although it is natural to associate an "ordered" material with
some sort of local structure, like a crystal, there are also other
order parameters that one may define. In particular, phase
separation of two liquids may be considered as a phase transition
where the order parameter is associated with the local mixing of
the fluids. Phase separation of polymer melts \cite{flory}, for
example, depends on the relation between the entropy gain of the
mixture versus the energetic advantage of the separated state. It
is well known \cite{polymerblends} that some systems of polymer
melts undergo phase separation when the temperature
\emph{increases}, a phenomena that, in some sense, is analogous
 to inverse melting (the "ordered", separated state is
thermodynamically stable only above some temperature). In the
Flory-Huggins \cite{flory} \cite{Huggins} theory of polymer melts,
the free energy contains a temperature dependent interaction term.
This implies that, to some extent, part of the internal entropy
associated with the possible conformations of a single polymer is
"absorbed" into the interaction term to yield an effective,
temperature-dependent interaction. The possible generalizations of
this Flory-Huggins procedure for inverse melting is discussed in
section VII.

\subsection{Kinetics of inverse freezing}

In any case of a first order transition between a liquid and a
crystalline phase, the system freezes into a glassy state if the
cooling is fast enough. The general kinetic description of this
phenomenon is based on the distinction between the nucleation rate
in a supercooled liquid (namely, the rate of creation of stable
nuclei of the crystalline phase) and the growth rate of a crystal.
Both processes are thermally activated and their rates admit
maxima between the melting temperature and $T=0$. At the melting
temperature, the bulk free energy associated with the two phases
is the same, and there is no driving force towards nucleation,
while as the temperature approaches zero, the kinetics of the
system halts due to the divergence of viscosity [the rate of any
thermally activated process depends on $exp(-\frac{\Delta E}{k_B
T})$]. There is a difference, however, between the locations of
these maxima, and in general one needs a lower temperature to get
a reasonable nucleation rate, since there is a minimal size for a
nucleus to be energetically favorable (the surface tension makes
small nuclei thermodynamically unstable even below the melting
temperature).

A good glass former liquid is associated, though, with diminishing
overlap between the nucleation and the crystal growth zone, i.e., at
temperatures just below the melting point there is no nucleation,
while at lower temperatures, nucleation actually takes place but the
crystal seeds could not grow as the viscosity diverges.

This picture yields a simple plausibility argument for inverse
freezing, i.e., for a liquid that forms glass as it absorbs heat.
Here, there is no decreasing kinetics as the temperature
\emph{increases}
 away from the melting point. Accordingly, any material that
undergoes inverse melting is, generally, a very bad glass former.
Unless some weird situation takes place, it is not plausible to
get a glassy state of matter as a result of fast heating of a
liquid. Accordingly, we suggest that inverse freezing appears,
generically, only in systems with quenched disorder or, at least,
if the glassy state is a true thermodynamic equilibrium state of
the system.

\section{Examples of inverse melting and inverse freezing}

Let us mention briefly  some examples of systems displaying inverse
melting which have been reported in the published literature,
classify them as first or second order, type I or II, and attempt to
explain their driving mechanisms shortly by the different sources of
the entropy and volumes in the different phases involved. It should
be stressed that the following list is by no means complete: a lot
of literature is devoted to the glass-crystal transition under the
name "reentrent" \cite{reentrant}, while the discovery of new
systems is reported \cite{new}.

\textbf{Helium isotopes $He^3$ and $He^4$:}  Both isotopes display
first order transition curves that qualitatively resemble the
neighborhood of point D in figure (\ref{fig1}), i.e. inverse
melting of negative slope (type I) \cite{helium}.  For both
isotopes the inverse melting happens at high pressures (about
25-30 bar) and, of course, at low temperature (less than 1K).
There is, however, a difference in the character of the solid and
the liquid phase. For $He^4$ a superfluid liquid  becomes an hcp
crystal upon heating (clearly the entropic gain here involves
longitudinal phonons). For $He^3$, on the other hand, normal (i.e.
non superfluid) liquid becomes a bcc crystal. This has to do with
nuclear spin degrees of freedom  that are relatively free to
reorient independently in the crystal, thereby increasing its
entropy relative to the liquid.

\textbf{Metallic alloys:} Inverse melting transformations have also
been found in a number of binary alloys based on the early
transition metals Ti, Nb, Zr, and Ta with later transition metals
from groups V and VI. In inverse melting of alloys, a metastable
supersaturated crystallic alloy transforms polymorphously to an
amorphous state near the glass transition temperature upon heating
\cite{metallicalloy}. For example, metastable bcc $\beta$-TiCr
phases with Cr contents between $40\%$ and $65\%$ which were
prepared by mechanical alloying of elemental powder blends showed a
polymorphous transformation of the bcc alloy into an amorphous
phase. Furthermore, it was reported that this transition is
reversible, such that the alloy can be switched back and forth
between the amorphous and the bcc crystalline phase by application
of alternating annealing steps at $800^{0}C$ and $600^{0}C$. From
these results and also numerical thermodynamic calculations it was
obtained that at those temperatures, a thermodynamic driving force
must exist for the amorphization such that the free energy of the
amorphous phase is lower than that of the bcc alloy for those
configurations. The occurrence of inverse melting originates from a
pronounced short-range ordering of the amorphous phase upon
undercooling, which stabilizes the amorphous phase with respect to
the bcc. Thus, although the crystal is much more topologically,
long-range, ordered than the amorphous, the amorphous phase admits
much more chemical short range order and therefore is of lower
entropy.

\textbf{Liquid crystals:} An analogue to inverse melting, whose
driving force is similar to the metallic alloys is provided by
liquid crystals.  It was shown that a first order boundary between
smectic-A and nematic phases of 4-cyano-4'-octyloxybiphenyl
(called 8OCB) looks very much like the portion A $\to$ B  $\to$ C
$\to$ D of Figure (\ref{fig1})  \cite{Johari} \cite{Cladis}. The
8OCB liquid crystal molecule contains both a polar and a nonpolar
part, as lipid bilayers. It consists of a flexible n-octane chain
attached to a relatively rigid 4-cyano-biphenyloxy group. It was
shown (by optical microscopy) that upon heating at a constant
pressure the nematic phase transforms into a smectic-A phase, and
on cooling again, it reversibly transforms back to the nematic
phase. The nematic low temperature state possesses just molecular
orientational order, while the smectic-A high temperature phase
possesses both orientational and partial translational order. Long
range attractive electrostatic forces stabilize layering, while
the short range repulsive interactions stabilize the nematic phase
at low temperatures.  Thus, in this material many internal degrees
of freedom are coupled to the orientational and positional order
to produce an inverse melting analogue. In addition, "re-entrant"
nematic $\longleftrightarrow$ smectic-A transformations were
observed in binary mixtures of analogous molecules. The
thermodynamics of the binary mixtures may involve also an
alloy-like chemical short range ordering in the nematic phase. It
was proposed  \cite{Johari}, that tight but mobile configurations
of associated molecular pairs reduce the entropy of that phase.

\textbf{Ferroelectricity in Rochelle Salt} Rochelle salt
($NaKC_{4}H_{4}O_{6}\cdot 4H_{2}O$, double sodium potassium
tartrate tetrahydrate) is a ferroelectric material exhibiting two
Curie points: one at $-18^{0}°C$ and the other at $+24^{0}°C$
\cite{rochelle}. This material is ferroelectric with a monoclinic
point group 2 and, in its  non-ferroelectric region, its structure
belongs to the orthorhombic point group 222. The higher Curie
point is similar to regular ferroelectric transition, however, the
lower point - the point where the spontaneous polarization is
lost, and the system becomes paraelectric (disordered) - is not
trivial, since the crystalline structure above the upper Curie
point and  below the lower Curie point are \emph{the same}. This
time the inverted transition is second order in type. Both the
higher and the lower Curie point go up in temperature at higher
pressure, i.e., $dP/dT>0$. In the next section the theoretical
explanations for this behavior are presented.

\textbf{Water} The liquid-liquid transition theory for
polyamorphous materials, i.e. materials that can have more than
one amorphous form, predicts an    inverse freezing transition
even for the most known liquid, water. In the hypothesized phase
diagram presented in \cite{stanley}, below a second critical point
with coordinates $T=220K$ and $P=100MPa$, the liquid phase
separates into two distinct liquid phases: a low density liquid
(LDL) phase at low pressures and a high density liquid (HDL) at
high pressures. Between these points water is a fluctuating
mixture of molecules whose local structures resemble the two
phases, LDL and HDL. The small region between $100MPa$ and
$150MPa$ and temperatures between $-50^{0}C$ to $-100^{0}C$
exhibits a range of inverse melting where the low density
amorphous becomes a low density liquid upon cooling. Although the
region of this hypothetic inverse freezing scenario is not
accessible experimentally, it is interesting to note the
possibility of an inverse transition even is the most familiar and
important liquid on earth.

\textbf{Magnetic films} An inverse transition effect is also found
in ultra-thin Fe films that are magnetized perpendicular to the film
plane \cite{Portmann}. The magnetization of these films is striped
domains with opposite perpendicular magnetization. From scanning
electron microscopy it was found that when the temperature is
increased, the low temperature stripe domain structure transforms
into a more symmetric, labyrinthine structure. However, at even
higher temperatures and before the loss of magnetic order, a
re-occurrence of the less symmetric stripe phase is found. The
mechanism driving this transition is topological defects such as
dislocations and disclinations. More specifically knee-bend and
bridge instabilities lead to the straightening of the labyrinthine
pattern when the temperature is increased. Thus the increase in
topological disorder drives the transition.

\textbf{Vortex lines in disordered high temperature superconductor:}
First order, type II transition from glassy to a crystalline state
was discovered in the  lattice formed by magnetic flux lines in a
high temperature superconductor $Bi_{2}Sr_{2}CaCu_{2}O_{8}$ (BSCCO)
\cite{pinning}.  The ordered hexagonal lattice has larger entropy
than the low temperature disordered phase. The explanation suggested
is that the transition from the lattice to the glass phase is driven
by pinning of the flux lines to impurities in the crystal at low
temperatures. The competition between thermal fluctuations and
pinning disorder leads to inverse melting near the critical point.
In this system, however, the intensive order parameter (bulk
magnetization) is lower in the crystalline phase, and the response
functions are higher, i.e., the disordered phase is stiffer than the
ordered phase.

\textbf{"Cold denaturation" of proteins:} Most of the proteins
denaturate, i.e., lose their biologically active, native state, at
high temperatures. Since a protein is a complex object with many
degrees of freedom, its denaturation transition resembles a "true"
first order transition in an infinite system \cite{Proteins}. In
contrast with the "regular" denaturation upon heating, the protein
ribonuclease A displays a reversible "inverse denaturation" upon
cooling (type II inverse melting) at high pressure (about 4kbarr).
This phenomenon may be explained on the basis of the internal
structure of the protein itself, as secondary or higher order
structures are lost upon denaturation. A different explanation
which has been proposed is the loss of  "low density water" as the
cause for cold denaturation \cite{ColdDenat}, this has been
modelled and found in agreement with the experimental data. In
addition to the study of ribonuclease A, cold denaturation at very
high pressures has also been observed in other biological systems
\cite{Biology}.

\textbf{ Colloidal systems- PMM1 sticky spheres:}  A simple model
system which was studied both theoretically and experimentally is a
collection of hard spheres in a given volume. Hard sphere particles
are increasingly caged by their neighbors as the density increases
and at a critical density, the system becomes non-ergodic or glassy.
The glass transition in that case depends only on the filling
fraction of the system and is independent of temperature, as the
thermal energy is negligible compared with the repulsion.  The
addition of short-range inter-particle attraction (stickiness)
introduces new energy scale, and a corresponding temperature, into
the problem. It was shown that, as temperature decreases, the
attraction first 'melts' the "hard sphere" glass, thus causing an
inverse freezing transition, and then, upon further decrease of the
temperature, a second, qualitatively different, glassy state is
formed due to the attractive interactions. Experimentally
\cite{Pham2} the system consisted of a colloidal system of
sterically stabilized polymethylmethacrylate (PMMA) particles,
dispersed in cis-decalin, with short range attraction induced by
adding a non-adsorbing polymer, polystyrene. The polymer is excluded
from the region between the surfaces of two nearby particles, thus
leading to an excess osmotic pressure attracting the particles
together. From the behavior of the samples, it was found that the
line of structural arrest at the high-density end of the phase
diagram has a re-entrant phase. This has also been observed by MCT
calculations \cite{Pham}, MD simulations, and light scattering
experiments which all suggest that the qualitatively distinct kinds
of glasses are dominated by repulsion and attraction, respectively.

\textbf{Polymeric systems:}

\textbf{(a) Poly (4-Methylpentene-1):}  A different inverse
melting material is the polymeric substance
poly(4-methylpentene-1) denoted more simply as P4MP1\cite{greer}.
This is a semi-crystalline one component polymeric system having a
crystalline component of nearly $60\%$. Below the glass transition
temperature (at around room temperature and atmospheric pressure),
the crystal density of the polymer is lower than the amorphous
phase. Therefore, on compression, the initially crystalline
tetragonal phase loses order and becomes amorphous above a
threshold value of 2kbar. This transformation is exothermic in
nature, thus suggesting that the amorphous phase has lower entropy
than the crystalline tetragonal phase. Indeed, a disordering on
cooling of the crystalline phase, that is, inverse melting and
crystallization on heating were observed. These structural changes
have also been confirmed by other experimental methods. It was
observed that the melting curve in the T-P plane possesses a
maximum of the type shown in figure (\ref{fig1}) by point B and
its neighborhood, i.e. the slope of the inverse melting curve is
positive (Type II).  This 'solid state amorphization' is in
agreement with the unusual density relationship below the glass
transition temperature of the polymer. The mechanism for the
inverted transition is the larger amount of conformations of
backbone and side-groups of the polymer in the crystal, which are
due to its more open structure, and this contributes to its
overall higher entropy.

\begin{figure}
\includegraphics[width=7.7cm]{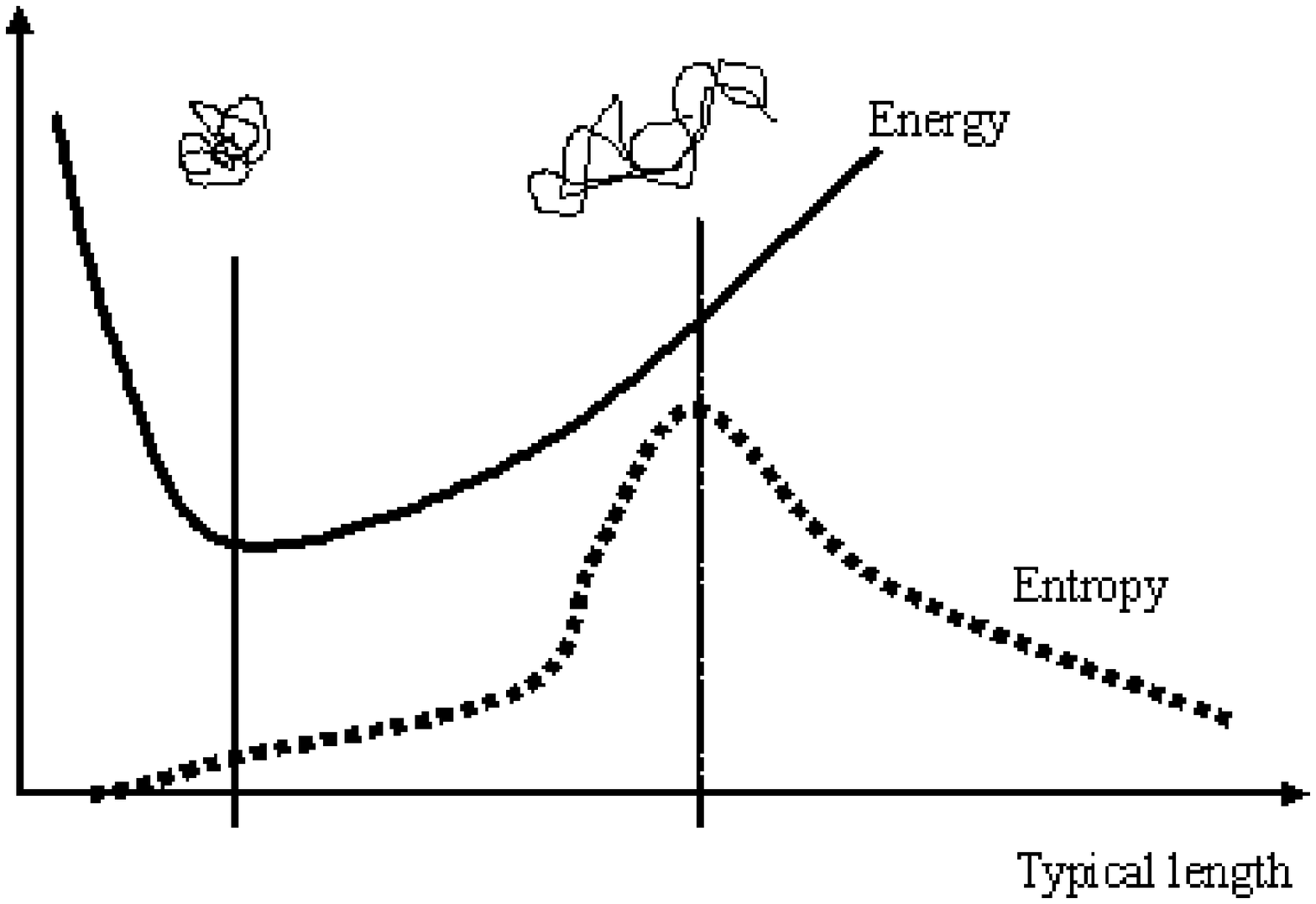}\\
\caption{A sketch of the energy and entropy dependence on the
linear size of a Methyl Cellulose polymer in water. The folded
conformation costs less energy due to more favorable interactions
between hydrophobic sequences along a single chain, but are less
entropic as water molecules have less freedom around the
hydrophobic constituents of the chain. The unfolded conformation
admits much more microscopic configurations. The interaction with
other polymers in the solution is suppressed in the folded state
and therefore it constitutes the liquid.} \label{fig2}
\end{figure}

 \textbf{(b) Methyl Cellulose:} An interesting example in polymeric
systems for inverse glass transition is the reversible
thermogelation of Methyl Cellulose solution in water
\cite{Chevillard}. When a (soft and transparent) solution of
Methyl Cellulose is heated (above $55^\circ$C, for a 5 gr/liter
solution) it turns into a white, turbid and mechanically strong
gel. This transition is reversible, and upon subsequent cooling,
the polymer is redissolved again. In its high temperature phase,
Methyl Cellulose gel exhibits, like many other gels \cite{gel},
glassy features. In this case, the folded conformation is favored
energetically while its unfolded conformation is favored
entropically [See figure (\ref{fig2})]. The entropy growth of the
open conformation may be related to the number of possible
microscopic configurations of the polymer itself, but it may be
attributed also to the spatial arrangement of the water molecules
in its vicinity, similar to the process suggested before for
protein denaturation. The mechanism proposed also for other
systems displaying inverse transitions due to the hydrophobic
effect \cite{hydrophob} is as follows: In the liquid state the
water molecules are kept in a highly constrained 'cage like'
structures formed by the hydrophobic constituents which move
around in the solution.  However, as the gel is formed, and the
hydrophobic segments cluster together to form cross-links, these
cages are opened, and the water molecules move freely around the
network. As a consequence, the number of possible configurations
and the entropy of the water molecules (which highly determines
the entropy of the whole system consisting of $99\%$ water) is low
in the liquid phase and increases when hydrophobic aggregates
cluster together and form a gel \cite{haque}. The main cause for
inverse glass transition is that the "open" high entropy
conformations of the polymer are also the \emph{interacting}
structures, as they allow for the formation of hydrophobic links
with other polymers in the solution, a process that leads to
gelation.

\textbf{(c) Other Polymers:} Aqueous solutions of the triblock
copolymer PEO-PPO-PEO (PPO - polypropylene oxide, PEO -
polyethylene oxide) also show inverse melting
behavior\cite{Mortensen}. Similar to methyl Cellulose, due to the
entropic mechanism, the PPO block is hydrophobic at high
temperatures and hydrophilic at low temperatures. Above a certain
concentration and a specific temperature, the Gaussian chains of
these polymers form micelles. At even higher temperatures it is
found that the micellar liquid transforms into a stable cubic
(bcc) crystal. In contrast to the methyl Cellulose, the transition
is to an ordered solid since the hydrophobic sequences are
deposited at ordered positions along the chain. The entropy change
due to crystallization may be small compared to the entropy change
of molecular origin and this is the assumed mechanism for the
inverse melting transition.
\begin{table}
  \includegraphics[width=7.7cm]{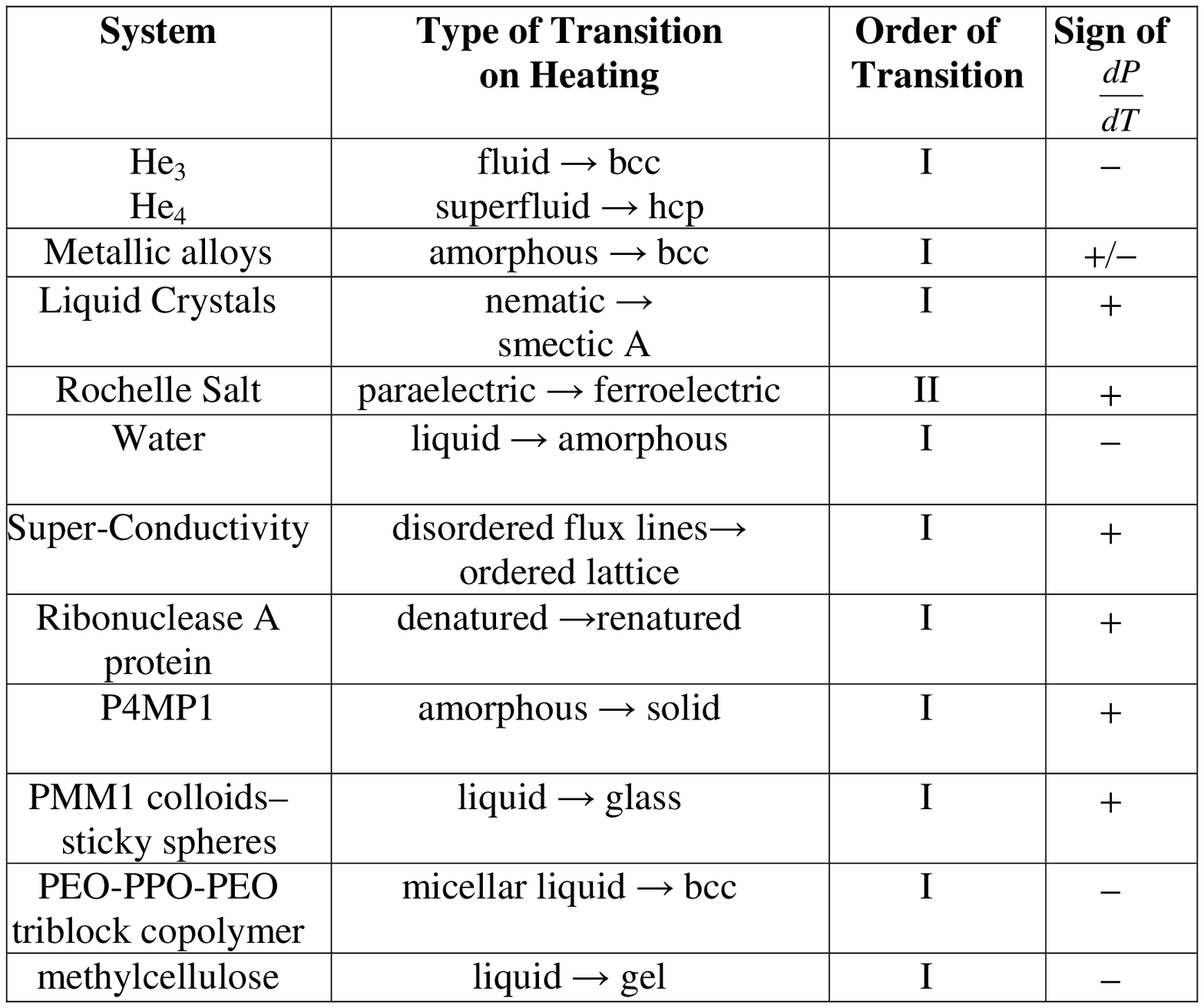}\\
  \caption{A summary of the different known physical systems
exhibiting inverse melting and the transition characteristics.}
\label{table1}
\end{table}

\section{Theoretical Modelling}

As mentioned above, the Flory-Huggins theory of phase separation
in polymer melts may yield phase separation as temperature
increases. In this theory, however, a temperature dependent
interaction is included. In this section we try to review several
"first principle" models that exist in the literature. In these
models, the ground state energy and the excitation spectrum are
temperature independent, and the inverse transition is attained by
applying the thermodynamic consideration to the given spectrum.

\textbf{Rochelle salt:} Perhaps the first theoretical
considerations that dealt with inverse melting appeared in the
context of the two Curie points for the ferroelectric phase  of
Rochelle salt \cite{mitsui}. The model may be presented in a form
of "quantum" pseudo-spin  model, where the Hamiltonian is
non-diagonal in the $z$ direction \cite{blinc} and two sublattices
are defined with different local field and interactions. The
combined effect of thermal and quantum tunneling dominate the
system in some temperature range to yield a finite magnetization
in the $z$ direction.

\textbf{Random heteropolymer in a disordered medium:} In a recent
model presented by Shakhnovich et. al. \cite{randompoly}, a random
heteropolymer in a disordered medium is considered. Here, as in
the flux line crystallization problem \cite{pinning}, the low
energy, low entropy state of the polymer involves pinning by
quenched  randomness that corresponds to a wandering exponent
larger than the $1/2$ value associated with thermal wandering and
crystallization is avoided. At larger temperatures the impurity
pinning may be neglected and the thermally wandering polymers are
free to form a structure based on their mutual interactions. In
case of random heteropolymers, this structure is glassy, as
opposed to the crystalline structure obtained in the flux line
case.

\textbf{Extended gaussian core model:} In a recent work by Feeney,
Debenedetti and Stillinger \cite{feeney}, an extension of the
Gaussian core model has been presented as a model that includes
first order inverse melting transitions. The particles of this
model are point particles that interact via a Gaussian repulsive
potential, and in the extended model any single particle may be in
one of two internal states, where the ground state is non
degenerate and the excited state admits high degeneracy. If the
interaction range of the excited states is shorter than the
interaction range of a particle in the ground state, the effective
density of particles decreases as temperature is increased. This
leads to a type II (water like) inverse melting scenario, since
the molar volume of the solid is larger than that of the liquid.
On the other hand, if in the extended model the interaction range
for the ground state is shorter, heating induces larger effective
density that yields type I inverse melting scenario.

\textbf{Extended Blume-Capel model:}  The spin model presented
below contains the basic ingredients of the extended gaussian core
model in a spin system. Its advantage relays on the simplicity of
modeling and solutions, and it gives a unifying framework to
analyze both inverse melting and inverse freezing of type I, type
II, first and second order. Although this model is not directly
related to any of the systems presented above, it  may yield
various general insights into the inverse transitions as shown in
the next sections.

\section{Spin model for inverse melting: the ordered case}

\subsection{a model for inverse melting of type I }

It has already been explained that, in order for inverse melting
of type I to occur, the more frozen, interacting, state has to be
of higher entropy, i.e. to have more internal configurations, than
the liquid noninteracting state. In order to model this phenomenon
in a most simple way one should look for the simplest model that
incorporates  all these features.  Here we use  a modified version
of the Blume-Capel model \cite{Blume}. The fundamental
constituents are spin one particles, and there are two competing
interactions: an exchange interaction that  lowers the energy of
the $\pm 1$ (interacting) states, and a "lattice field" that
favors the  "zero" (noninteracting) state. For $N$ interacting
spins the Blume-Capel (BC) Hamiltonian takes the form:
\begin{equation}
\label{eq:4} H=-J\sum_{<i,j>} S_{i}S_{j}+D\sum_{i=1}^N S_{i}^2-
h\sum_{i=1}^N S_{i}
\end{equation}
where the spin variables are allowed to assume the values
$S_i=0,\pm 1$. The summation over $<i,j>$ is over any interacting
pair once and $h$  is the magnetic field applied. The magnetic
field term that breaks the up down symmetry of the spins has no
direct relevance to the inverse melting and is included here only
for completeness of the discussion and  for susceptibility
calculations. Nevertheless, the phase diagrams below, will be
plotted for $h=0$.

For positive $D$ the noninteracting state of a single spin is
lowered in energy  than the interacting state. For $D>qJ$ (where
$q$ is the number of interacting particles, or "nearest
neighbors", of the model), the ground state of the Hamiltonian is
the "folded" state, where all spins are zero, i.e., the system is
in its noninteracting phase. For $D < qJ$ it is favorable for the
system to be in its interacting phase, and at the ground state all
spins are either at the $+1$ or at the $-1$ states. At zero
temperature this implies a transition, upon increasing $D$, from
the ferromagnetic state to the paramagnetic one, and spontaneous
breakdown of the up-down symmetry in the interacting phase. Thus,
this model already includes one basic ingredient of the inverse
melting scheme, namely, the energetic preference of the non
interacting state.

In order to explain the second constituent essential for our model
let us use the Methyl Cellulose analogy [see Figure (\ref{fig2})]
as an example. If the zero spin state of the BC model represents
schematically the compact non-interacting polymer coil, the
stretched polymer (interacting with its neighbors) is represented
by spin $\pm 1$. Clearly there are many possible spatial
configurations in which two polymers may attach to each other, and
correspondingly many degenerate, or almost degenerate, frozen
configurations of the gel; in our schematic model this is
represented by the degeneracy between plus and minus states.

The new ingredient that should be added to the classical BC model in
order to yield inverse melting is the entropic advantage  of the
interacting states. As a first approximation let the $0$ spin state
to be  k-fold degenerate, and the $\pm 1$ states to be are l-fold
degenerate where $r=l/k \geq 1$ is the degeneracy ratio that
dictates the entropic advantage. It turns out that all the results
presented here are independent of the absolute degeneracies $k$ and
$l$, and depend only on their ratio $r$. The parameter $r$
represents, of course, the more configurations available for a
polymer in its opened (interacting) states relative to the number of
configurations it can obtain in the closed (noninteracting) coil.

The Blume-Capel model, as well as its modification presented here
may be easily solved in its infinite range limit, i.e., where
there is no spatial structure and any pair of spins interact with
each other. In order to keep the effective field finite  one
replaces the exchange factor in the Hamiltonian $J$  by $J/N$.
Using standard gaussian integral techniques one finds an
expression for the free energy per spin in the infinite range
limit:
\begin{equation} \label{diff}
 \beta f \equiv \beta F/N  =\frac{\beta J M^2}{2}-
ln \{1+2 \ r \ cosh[ \beta (J M+h)] e^{-\beta D} \}
\end{equation}
\noindent where $M$ is the order parameter of the system
(magnetization per spin), $M \equiv \langle \frac{1}{N}
\sum_{i=1}^N S_i \rangle$. The phase transition curves are
obtained numerically by solving for the minimum of Eq.
(\ref{diff}) with respect to $M$, namely, the equation
\begin{eqnarray} \label{minm}
 M=\frac{2r \sinh [\beta (J M+h)] }{e^{\beta D}+2 r \cosh [\beta (J
M+h) ]}
\end{eqnarray}
should be solved self consistently.

\begin{figure}
  \includegraphics[width=7.7cm]{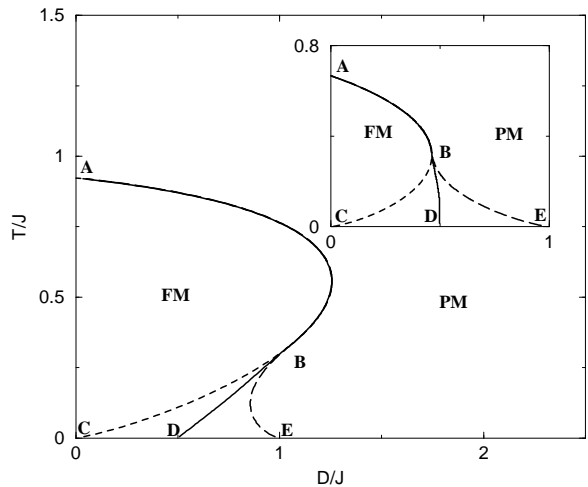}\\
  \caption{Phase diagram and spinodal lines for the ordered
  model Eq. (\ref{diff}) in the $D-T$ plane
  for $r=1$  (Blume-Capel model, inset) and for $r=6$.
  The value of $r=6$ has been chosen in order for the effect to be more
  pronounced, but inverse melting is seen as soon as $r>1$. As mentioned in
  the text, the effect of introducing a magnetic
field will not be considered and all phase diagrams will be
plotted for $h=0$. } \label{fig3}
\end{figure}

Scaling the temperature and $D$ with the interaction strength $J$,
the phase diagram is shown in Figure (\ref{fig3}). In the inset,
results are presented for the original Blume-Capel model (i.e.,
the $r=1$ case): the line AB is a second order regular transition
line, above it is a paramagnetic ($M=0$) phase and below it the
system is Ferromagnetic ($M \neq 0$). Below the tricritical point
(B) the phase transition is first order, and the three lines
plotted are: the spinodal line of the ferromagnetic phase BE
(above this line the $M\neq0$ solution ceases to exist), the
spinodal line of the paramagnetic phase BC (below this line $M=0$
is not a minimum of the free energy) and the first order
transition line BD. Along BD the free energy of the paramagnetic
phase is equal to that of the ferromagnetic state. Clearly, the
Blume-Capel model displays no inverse melting: an increase of the
temperature induces smaller order parameter.

The situation is different as $r$ increases, as emphasized by the
main part of Figure (\ref{fig3}). The same phase diagram is
presented, but now $r=6$, so the interacting states have larger
entropy. The ferromagnetic phase now covers a larger area of the
phase diagram, a fact that reflects its entropic advantage. The
tricritical point is shifted to the left, relative to the point of
infinite slope, leaving a region of second order inverse melting,
and the orientation of the BD line also changes, establishing the
possibility of first and second order inverse melting. Note that the
$r=6$ transition  lines converge to the $r=1$ lines as  $T \to 0$,
since  the entropy has no effect on the free energy at that limit.

The value of $r=6$ was chosen only for illustration. In fact, as
soon as $r$ gets larger than 1, inverse melting of first order is
observed. For $r=1$, i.e. the original Blume-Capel model, the
tricritical point is placed a bit higher then the point of
infinite slope and the BD line curves to the right. However, as
$r$ increases a bit, a small portion of the BD line obtains
negative curvature, thus inserting a small region of first order
inverse melting. However, the general trend of the first order
transition line BD is still to the right. The tricritical point
begins to move downwards through the AB line and continues to do
so on a further increase of $r$. It crosses the point of infinite
slope for $r\sim1.1204$ and thus for larger values of $r$, first
and second order regions of inverse melting occur as the
tricritical point continues to move downwards on the melting curve
to below the point of infinite slope. All in all, it seems that
the original Blume-Capel, i.e. for  $r = 1$ is exactly "marginal"
in the sense of inverse melting.

\begin{figure}
  \includegraphics[width=7.7cm]{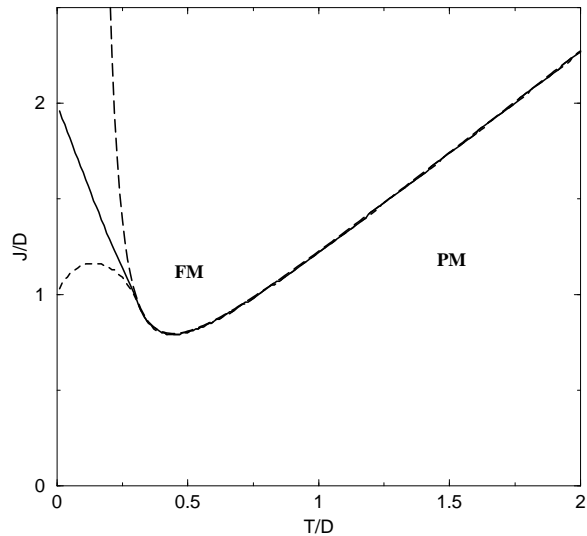}\\
  \caption{Phase diagram and spinodal lines for the ordered
  model Eq. (\ref{diff}) in the $T-J(P)$ plane
  for $r=6$. First and second order inverse melting of the first type
  are observed, i.e. the slope of the curve is negative. } \label{fig4}
\end{figure}

To allow qualitative comparison of our cartoon model with
experimental results, the appropriate parameters should be
identified. There are three parameters in the model as it stands:
$D$ represents the energetic advantage of the noninteracting
state, $r$ (if larger than 1) is the entropic gain of the
interacting state, and $J$ is the strength of the interaction. In
most of the physical systems that display inverse melting, the
controlled external parameter is the strength of the interaction:
pressure (for $He^3$ and $He^4$) or concentration of the
interacting objects (for polymeric and colloidal systems and
Rochelle salt - Ammonium Rochelle salt mixtures). As long as the
only effect of the pressure is to increase the strength of the
effective interaction among constituents, it may be modelled by
changing $J$. The resulting phase diagram should be compared,
though, with the $T-J$ plot of our model presented in Figure
(\ref{fig4})  and shows type I, i.e. negative slope, inverse
melting. The decrease of the transition temperature with the
increase of interaction strength (pressure) is physically
intuitive, as larger interaction favors energetically the
ferromagnetic phase. As discussed previously, the slope of the
first order transition line in the temperature-pressure plane is
required by the corresponding Clausius-Clapeyron equation
(\ref{eq:Clausius}).

Inverse melting obtained by this model is type I, $V^{(s)}<
V^{(l)}$, and one expects a negative slope of the transition line as
was also shown in portion AD of figure (\ref{fig1}). In real
magnetic or electric system the intensive-extensive pairs [magnetic
field-magnetization ($\textbf{H} \cdot d\textbf{M}$) or electric
field-polarization ($\textbf{E}\cdot d\textbf{P}$)], appear in the
free energy function with inverse sign relative to $PdV$. If the
order parameter vanishes, or takes smaller values, in the "liquid"
(disordered) phase, this implies also negative slope of the first
order transition line in the temperature-external field plane.

\subsection{a model for inverse melting of type II }

\begin{figure}
  \includegraphics[width=7.7cm]{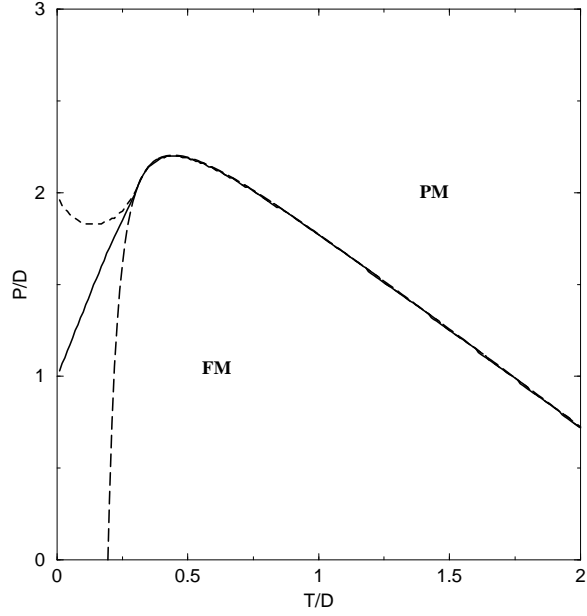}\\
  \caption{Phase diagram and spinodal lines for the ordered
  model Eq. (\ref{eq:second}) of the second type in the $T-P$ plane
  for $r=6$. The slope of the inverse melting curve is positive this
  time.
  } \label{fig7}
\end{figure}

The modelling of  type II inverse melting,  where the volume of
the liquid is lower than that of the crystal is possible by slight
modification of  the Hamiltonian. Following \cite{ColdDenat} we
add an additional energetic 'cost' to the interacting states of
the hamiltonian, thereby reducing the net gain in the free energy
obtained for a large $r$. Thus the frozen interacting state is
also of higher volume than the noninteracting one and its energy
increases. The Hamiltonian becomes
\begin{equation}
\label{eq:second} H=-J\sum_{<i,j>} S_{i}S_{j}+D\sum_{i=1}^N
S_{i}^2+P \delta V \sum_{<i,j>} S_{i}S_{j},
\end{equation}
where $\delta V$ is the access volume of the "open", interacting
configurations. Plotting the locus of the phase transition, this
time as a function of pressure, yields the phase diagram shown in
figure (\ref{fig7}). The slope of the $P(T)$ diagram is shown
positive. This is analogous to raising the energetic cost of the
interacting state to a state where $J_{eff}=J-P \delta V$.
Therefore, for higher values of  $P$, higher temperatures are
needed to induce the inverted transition.

\subsection{Response functions}

To complete the picture let us calculate the values of some
thermodynamic quantities that characterize the transitions in the
first version of the model. The heat capacity, given by
\begin{equation} \label{heatordered}
C_{H}= 2 r k_B \beta^{2}\exp (\beta D)\cdot \frac{ 2 r J ^2 M^2 -2 D
J M \sinh [\beta (J M+h)] + (D^2 +J^2 M^2)
  \cosh [\beta (J M+h)]}{\left[\exp (\beta D) +
  2 r \cosh [\beta (J M+h)] \right ]^2 }
\end{equation}

\begin{figure}
  \includegraphics[width=7.7cm]{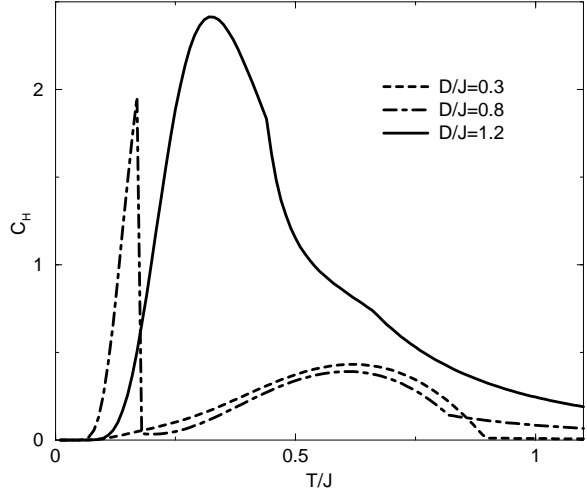}\\
  \caption{Heat capacity as a function of the scaled temperature $T/J$ for the
  ordered model, Eq. (\ref{diff}), for different values of $D/J$.  First order
  inverse melting shows a jumps in the heat capacity while second order
  transitions show a discontinuity in the derivative of the heat capacity. } \label{fig5}
\end{figure}

In figure (\ref{fig5}) the heat capacity as a function of
temperature is shown for  different values of $D/J$ (i.e., along
vertical sections of the phase diagram (\ref{fig3}). For small
$D/J$, where there is no inverted phase transition (demonstrated
in the figure by $D/J=0.3$), the heat capacity shows only a
monotonic increase to a maximum followed by a decrease as expected
by thermal changes. In the point of second order (normal) melting,
the slope changes suddenly. Since the model is globally coupled,
there is no meaning to the correlation length and domain size and
therefore, the diverging heat capacity associated with second
order transitions is avoided. For higher values of $D/J$ (shown
for $D/J=0.8$),  as the first order inverse transition sets in,
there is a discontinuity of the heat capacity. Following this is
an increase, then a decrease, of the heat capacity with the
temperature and again an abrupt change of the slope when the
second order melting occurs. Second order inverse melting, (as
obtained for $D/J=1.2$) also shows an abrupt change in the
derivative of the specific heat at the transitions (not easily
seen in the figure  at $T=0.5$ and $T=0.75$). For higher values of
$D/J$ there are no phase transitions so the heat capacity is a
monotonic function.

\begin{figure}
  \includegraphics[width=7.7cm]{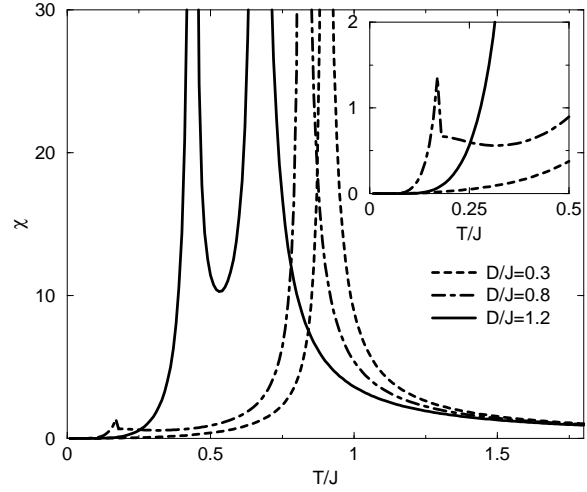}\\
  \caption{Susceptibility as a function of the scaled temperature $T/J$ for the
  ordered model, Eq. (\ref{diff}), for different values of $D/J$. First order
  inverse melting shows a discontinuity
  in the susceptibility while second order transitions show divergences.}
   \label{fig6}
\end{figure}

The susceptibility as a function of temperature is  given by
\begin{equation} \label{susceptordered}
 \chi= \left( \frac{2r \sinh^{2} (\beta J M) ^{2}}{\beta M ^{2}[2r+ \exp (\beta D)
 \cosh (\beta J M)]}-J \right)^{-1}
\end{equation}
and several vertical cuts are shown in Figure (\ref{fig6}). First
order inverse melting yields a small discontinuity in the
susceptibility as seen more clearly in the inset for $D/J=0.8$.
However, all second order transitions in the system, including the
inverted ones give diverging values of the susceptibility at the
transitions, as expected.

\section{spin model for inverse freezing}
\subsection{Model system and the replica trick}

As already explained in the introduction, inverse freezing is  the
(reversible) appearance of glassy features in a system upon
raising the temperature. This  may be incorporated in our spin
model by introducing random coupling $J_{ij}$, as in the standard
spin-glass models \cite{binder}.  The random-exchange
generalization of the Hamiltonian (\ref{eq:4}) is:
\begin{eqnarray}
\label{eq:13} H=\sum_{<i,j>} J_{ij}S_{i}S_{j}+D\sum_{i=1}^N
S_{i}^2- h\sum_{i=1}^N S_{i}
\end{eqnarray}
where the exchange interaction between the $i$ and the $j$ spin is
taken at random from some predetermined distribution. Following
the paradigmatic Sherrington-Kirkpatrick (SK) analysis
\cite{binder} of the infinite range  spin glass, we assume
gaussian distribution of the exchange term with zero mean:
\begin{eqnarray}
\label{eq:14} P(J_{ij})=\sqrt{\frac{N}{2\pi
J^{2}}}\exp-[\frac{N(J_{ij}- \frac{J_{0}}{N})^{2}}{2J^{2}}],
\end{eqnarray}
where $\frac{J_{0}}{N}$ is the mean of the distribution and
$\frac{J}{\sqrt{N}}$ is its width.   The replica trick is then
implemented to get the free energy at the large $N$ limit.

The case $r=1$, namely the random  exchange version of the
Blume-Capel model,  was first introduced and discussed by Ghatak and
Sherrington  (GS) \cite{Ghatak} who used symmetric replica to obtain
the relevant phase diagram. The GS solution seems to display inverse
freezing very pronouncedly even for the $r=1$ case, but more
detailed analysis by da Costa et. al. \cite{Costa} revealed errors
in the numerical stability  analysis  of GS. They discovered that
the glass order-parameter takes nonzero values (with a variety of
stability features) in the area below the GS transition line, and
the temperature dependence is monotonic. In their corrected solution
only a very tiny "inverse melting" region exists. Recently, the full
replica symmetry breaking analysis has been implemented for the GS
model \cite{crisanti}, and the results admit no inverse glass
transition. Here we present a replica symmetric analysis of the same
hamiltonian where the interacting states are highly degenerate,
i.e., $r>1$. Following \cite{Costa}, we obtain the phase transition
and the spinodal lines, and the results support, again, both first
and second order inverse glass transition. The generalization of the
replica symmetry breaking technique to both the Blume-Capel and the
Blume-Emery-Griffith models has been carried out by \cite{Arenzon}.

The replica technique \cite{Edwards} relies on the identity
\begin{equation}
\overline{ln[Z]}=lim_{n \rightarrow 0}\frac{1}{n}(\overline{Z^n}
-1), \end{equation} where $Z$ is the partition function of the
system and $Z^n$ is interpreted as the partition function of an
n-fold replicated system $S_i \rightarrow S_{ia}, a=1...n$. The
average free energy per spin may be computed using
\begin{equation}
\beta f =-lim_{n\rightarrow 0}\frac{1}{N n}(\overline{Z^n}-1).
\end{equation}
The disorder average is taken for $Z^n$ using the  Gaussian
distribution (\ref{eq:14}) and yields:
\begin{eqnarray}
\label{eq:15} \overline {Z^{n}} = Tr_{\{S_{i,a}\}} \exp \left[
\frac{\beta^2 J^2}{2N} \sum_{a>b} \left( \sum_iS_{ia} S_{ib}
\right)^2
 +\frac{\beta^2 J^2}{4N}\sum_{a}\left( \sum_i S_{ia}^2 \right)^2
 -\beta D
\sum_{a} \sum_{i} S_{ia} ^2  \right. \\  \nonumber  \left.  +
\frac{\beta J_{0}}{2N}\sum_{a} \left( \sum_{i}S_{ia}\right)^2 +
\beta h \sum_{a} \sum_{i}S_{ia} \right]
\end{eqnarray}
where $a,b=1...n$ are the replica indices. Implementing the
Hubbard-Stratanovitch identity yields the free energy per spin:
\begin{equation}
\label{eq:16} -\beta f = -\beta \frac{F}{N}= \lim _{n \rightarrow 0}
\frac{1}{n}\{-\frac{\beta^2 J^2}{2}\sum_{a>b}q_{ab}^2\
-\frac{\beta^2J^2}{4}\sum_{a}q_{aa}^2 -\frac{\beta
J_{0}}{2}\sum_{a}M_{a}^{2}+lnTr e^{ \hat{L}}\}
\end{equation}
where
\begin{equation}
\label{eq:17}
\hat{L}=\beta^2J^2\sum_{a>b}q_{ab}S_aS_b+\frac{\beta^2J^2}{2}\sum_{a}q_{aa}S_{a}^2
-\beta D \sum_{a}S_{a}^2 +\beta J_{0}\sum_{a}M_{a}S_{a}+\beta h
\sum_{a}S_{a}
\end{equation}
with $M_{a}$ the magnetization in each replica and $q_{aa}$,
$q_{ab}$, are the diagonal and the off diagonal entries of the
"order parameter matrix". All these  quantities are given
self-consistently by the saddle-point conditions:
\begin{equation}
\label{eq:18} M_{a}=\langle{S_a}\rangle \qquad
 q_{aa}=\langle{S_a ^2}\rangle \qquad
 q_{ab}=\langle{S_a S_b}\rangle
\end{equation}
as $\langle ... \rangle$ stands for thermal average over the
effective hamiltonian $\hat{L}$.

\subsection{replica symmetric solution}

In order to solve this model it is necessary to make assumptions on
the order parameter matrix elements  $q_{ab}$. In order to get a
general qualitative picture of the phase diagram of the system we
first make the simplest ansatz, which is symmetry with respect to
permutations of any pair of the replicas: $m_{a}=m, q_{aa}=p, \ \
\forall a$, and $q_{ab}=q, \ \  \forall a \neq b$. Using this
\emph{replica symmetric} assumption one obtains
\begin{equation}
\label {eq:10} -\beta f_{rs}=\frac{\beta^2
J^2}{4}(q^2-p^2)-\frac{\beta J_{0}}{2} M^{2}+
\frac{1}{\sqrt{2\pi}} \int_{-\infty}^{\infty} dz \exp (-\frac
{z^2}{2}) ln[1+2 \ r \ e^{\gamma}\cosh (\beta \tilde{H}(z))]
\end{equation}
with
\begin{eqnarray}
\label{eq:19} \tilde{H}(z)= J \sqrt{q} z +J_{0} M +h
\end{eqnarray}
and
\begin{eqnarray}
\label{eq:19} \gamma=\frac{\beta^2 J^2}{2}(p-q)-\beta D
\end{eqnarray}
Extremizing the free energy with respect to $q$, $p$ and $M$ one
gets the following set of  coupled equations:
\begin{equation}
\label{eq:20} q=\int_{-\infty}^{\infty}\frac{dz \ exp(-\frac
{z^2}{2})}{\sqrt{2\pi}}
 \left[ \frac {2 r  e^{\gamma}\sinh (\beta \tilde{H}(z))}
 {1+2 r e^{\gamma}\cosh (\beta \tilde{H}(z)) } \right] ^{2}
\end{equation}

\begin{equation}
\label{eq:21} p=\int_{-\infty}^{\infty} \frac{dz \ exp(-\frac
{z^2}{2})}{\sqrt{2\pi}}\frac {2 r e^{\gamma} \cosh(\beta
\tilde{H}(z))} {1+2 r e^{\gamma}\cosh(\beta \tilde{H}(z))}
\end{equation}

\begin{equation}
\label{eq:222} M=\int_{-\infty}^{\infty} \frac{dz \ exp(-\frac
{z^2}{2})}{\sqrt{2\pi}}\frac {2 r e^{\gamma} \sinh(\beta
\tilde{H}(z))} {1+2 r e^{\gamma}\cosh (\beta \tilde{H}(z))}
\end{equation}

\begin{figure}
  \includegraphics[width=7.7cm]{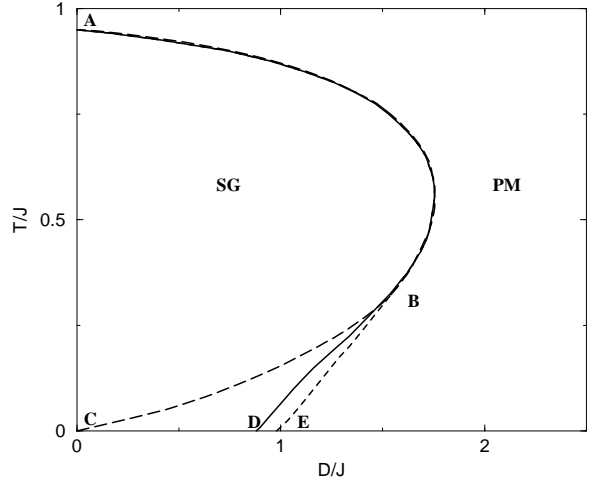}\\
  \caption{Phase diagram and spinodal lines for the disordered model
  in the D-T plane for a constant
  interaction $J$ for $r=6$.} \label{fig10}
\end{figure}

The  coupled equations (\ref{eq:20}),(\ref{eq:21}) and
(\ref{eq:222}) are  solved  numerically (with the possibility of
multiple solutions if more than one stable state exists). In the
limit where  $ h \rightarrow 0$ and $ J_{0} \rightarrow 0$ the
last equation vanishes. We will solve the equations in that limit
and then determine the location of the first order transition line
by comparison of the free energy values (plugging $q$ and $p$ into
(\ref{eq:10})), a procedure that ensures the continuity of the
free energy at the transition. The resulting phase diagram is
shown in Fig. (\ref{fig10}) for the case $r=6$, and displays all
the essential features that exist in the ordered model, including
inverse freezing of first and second order,  a tricritical point
and spinodal lines.

\begin{figure}
  \includegraphics[width=7.7cm]{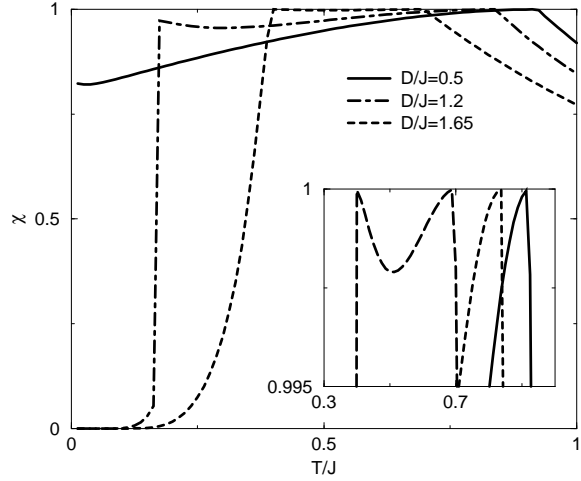}\\
  \caption{Susceptibility as a function of temperature for the
  glassy model for different values of $D/J$. First order inverse
  melting shows an abrupt jump in the susceptibility while cusps
  are seen for all of the spin glass second order transitions.
   } \label{glasssus}
\end{figure}

The susceptibility of the glassy model is given by
\begin{equation} \label{chi}
\chi=\beta (p-q)
\end{equation}
and shown in figure (\ref{glasssus}).  For low values of $D/J$
($D/J=0.3$) where there is no inverse freezing transition, the
susceptibility is a continuous function  and only shows a cusp in
the spin glass transition. However, as the inverse glass
transition sets in as a first order transition, the susceptibility
shows a discontinuity as shown for $D/J=1.2$ , and when it is
second order, the susceptibility consists of another cusp, similar
to the normal transition.

\begin{figure}
  \includegraphics[width=7.7cm]{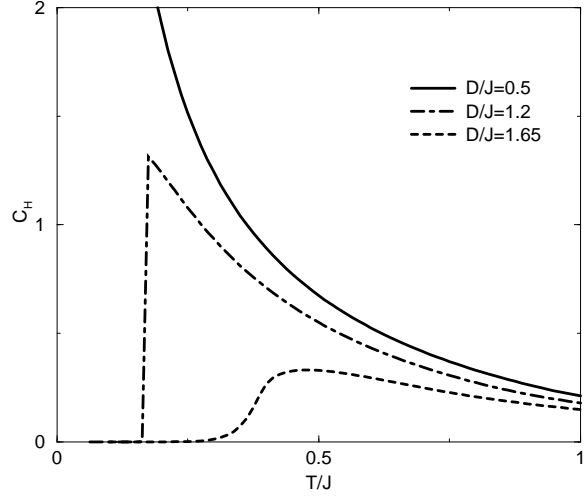}\\
  \caption{Heat capacity as a function of temperature for the
  glassy model for different values of $D/J$. First order
  inverse melting shows a discontinuity in the heat capacity while
   second order transitions do not show special features.} \label{glassch}
\end{figure}

The internal energy of the system is given by
\begin{equation}
\label{eq:22} U=\frac{\beta J^{2}}{2}\left( q^{2}-p^{2} \right)+D
p -\left( h+\frac{J_{0}M}{2} \right) M
\end{equation}
From the internal energy the heat capacity at a constant magnetic
field is calculated as:
\begin{eqnarray} \label{ch}
\label{eq:22} C_{H}=\frac{1}{2} k_B \beta ^{2} J ^{2}
(p^{2}-q^{2})
\end{eqnarray}
as seen from Figure (\ref{glassch}). Only first order transitions
show a feature of discontinuous heat capacity.

\subsection{Replica symmetry breaking}

It is well known that the replica symmetric solution suffers from
several problems, associated with ergodicity breaking in the
glassy state of matter, and that better and better solutions are
obtained by more steps in the replica symmetry breaking procedure
\cite{Mezard}. Here we briefly discuss the one step replica
symmetry breaking (1RSB) and comment about the full RSB in order
to clarify, in the next section, the basic features associated
with the degeneracy and inverse freezing.

One step RSB involves the division of the  off-diagonal elements
of the $n \times n$ matrix of $q_{ab}$ into $n/m$ blocks
containing $m$ replicas each. Different replicas in the same block
have overlap $q_{1}$ while those in different blocks have overlap
$q_{0}$.

Thus, the 1RSB free energy is given by

\begin{equation}\label{fr1s}
-\beta f_{1rsb}=\frac{\beta^2 J^2}{4}[m
(q_{0}^{2}-q_{1}^{2})+q_{1}^{2}-p^{2}] -\frac{\beta J_0 M^{2}}{2} \\
\nonumber   + \frac{1}{m \sqrt{2\pi}} \int_{-\infty}^{\infty} dz
\exp (-\frac {z^2}{2}) \\ \nonumber ln \left[ \frac{1}{\sqrt{2\pi}}
\int_{-\infty}^{\infty}d\zeta \exp (-\frac{\zeta ^{2}}{2})\cdot
\left\{ (1+2r e^{\gamma} \cosh [\beta
\tilde{H}(z,\zeta)])\right\}^{m} \right]
\end{equation}
where
\begin{eqnarray}
\label{eq:24} \gamma = \frac{\beta^2 J^2}{2} (p-q_{1})-\beta D
\end{eqnarray}
and
\begin{eqnarray}
\label{eq:25} \tilde{H}(z,\zeta)= J \sqrt{(q_{1}-q_{0})}\zeta + J
\sqrt{ q_{0}}z+J_0 M+h
\end{eqnarray}

As usual in spin glass theory, we have to maximize the free energy
as a function of $q_1$, $q_0$, $p $, $m$ and $M$, and the saddle
point equations are:

\begin{eqnarray} \label{q1-1rsb}
q_{1}=\int_{-\infty}^{\infty} \frac{dz \ exp(-\frac
{z^2}{2})}{\sqrt{2\pi}}\left\langle \left[ \frac {2 r e^{\gamma}
\sinh[\beta \tilde{H}(z,\zeta)]} {1+2 r e^{\gamma}\cosh[\beta
\tilde{H}(z,\zeta)]} \right] ^{2} \right\rangle_{A}
\end{eqnarray}

\begin{eqnarray} \label{q0-1rsb}
q_{0}=\int_{-\infty}^{\infty} \frac{dz \ exp(-\frac
{z^2}{2})}{\sqrt{2\pi}}  \left[ \left\langle \frac {2 r e^{\gamma}
\sinh[\beta \tilde{H}(z,\zeta)]} {1+2 r e^{\gamma}\cosh[\beta
\tilde{H}(z,\zeta)]}\right \rangle_{A} \right] ^{2}
\end{eqnarray}

\begin{equation} \label{p-1rsb}
p=\int_{-\infty}^{\infty} \frac{dz \ exp(-\frac
{z^2}{2})}{\sqrt{2\pi}} \left\langle \frac {2 r e^{\gamma}
\cosh[\beta \tilde{H}(z,\zeta)]} {1+2 r e^{\gamma}\cosh([\beta
\tilde{H}(z,\zeta)])}\right\rangle_{A}
\end{equation}

\begin{eqnarray} \label{M-1rsb}
M=\int_{-\infty}^{\infty} \frac{dz \ exp(-\frac
{z^2}{2})}{\sqrt{2\pi}}\left\langle \frac {2 r e^{\gamma}
\sinh[\beta \tilde{H}(z,\zeta)]} {1+2 r e^{\gamma}\cosh([\beta
\tilde{H}(z,\zeta)])}\right\rangle_{A}
\end{eqnarray}

The size of the inner blocks, $m$, satisfies

\begin{eqnarray} \label{m-1rsb}
\frac{\beta J ^{2}}{4}(q_{1}^{2}-q_{0}^{2}) +\frac{1}{\beta m^{2}}
\int_{-\infty}^{\infty} \frac{dz \ exp(-\frac
{z^2}{2})}{\sqrt{2\pi}}  \ln [ \frac{1}{\sqrt{2\pi}}
\int_{-\infty}^{\infty} d\zeta \exp (-\frac{\zeta ^{2}}{2}) A^{m} ]
\\ \nonumber - \frac{1}{\beta m}\int_{-\infty}^{\infty} \frac{dz \
exp(-\frac {z^2}{2})}{\sqrt{2\pi}} \left\langle ln [1+2 r
e^{\gamma}\cosh[\beta \tilde{H}(z,\zeta)] \right\rangle _{A} = 0
\end{eqnarray}

All these expressions need the definitions

\begin{eqnarray} \label{Az}
\label{eq:29} A(z,\zeta)= 1+2 r e^{\gamma}\cosh[\beta
\tilde{H}(z,\zeta)]
\end{eqnarray}
and
\begin{eqnarray} \label{XA}
\label{eq:30} \langle X \rangle _{A}=\frac{\int_{-\infty}^{\infty}
d\zeta \exp (-\frac{\zeta ^{2}}{2}) X
A^{m}}{\int_{-\infty}^{\infty} d\zeta \exp (-\frac{\zeta ^{2}}{2})
A^{m} }
\end{eqnarray}
where $q_{1}>q_{0}$ and all parameters are in the region $[0,1]$.
The numerical solutions are now obtained by either maximizing the
free energy or by solving the coupled system of  saddle point
equations (\ref{q1-1rsb}),(\ref{q0-1rsb}), (\ref{p-1rsb}),
(\ref{M-1rsb}) and (\ref{m-1rsb}). In the resulting phase diagram,
although the phase transition line is shifted a little to the
right the essential features of inverse melting remain the same.
The effect of replica symmetry breaking on $q_1$ is small, similar
to the SK model and other previously discussed models
\cite{Arenzon}.

This result is anticipated also from the following physical
intuition based on qualitative comparison to the ordered model.
The main difference, in the context of inverse melting, of the
disordered model from the ordered one, is the advantage of the
"frozen" state to be is less pronounced. Thus, although
frustration  yield less effective freezing then in the ordered
model, and therefore the free energy of the glass is higher, still
the interplay between the energy and entropy terms as a function
of the temperature remain the same. Therefore the qualitative
picture of the glassy system also, to any order of replica
symmetry breaking, should not be altered. In addition,
quantitatively, the ABC line that marks the spinodal of the $q=0$
phase is unaffected by the need to break the replica symmetry.
Since the points C,D,and E are $r$ - independent (at $T=0$ there
is no effect of entropy) one can choose always a large enough $r$
(like the one presented in the figure) in order to ensure that
some sort of inverse melting takes place, independent of the
degree of the symmetry breaking calculations.

Let us comment, now, on the full replica symmetry breaking (FRSB)
for this model. The FRSB involves an infinite process of blocks
within blocks, with an order parameter $q(x)$, $x \in [0,1]$
\cite{binder}. It is easy to see that, to any order in the RSB
process, the parameter $\gamma$ in (\ref{eq:24}) involves only the
\emph{inner} $q$, i.e., the order parameter associated with the
smallest blocks. As a result, the $\gamma_{FRSB}$ is:
\begin{equation}
\label{eq:24} \gamma = \frac {\beta^2 J^2}{2} [p-q(x=1)]-\beta D
\end{equation}
where $q(x=1)$ is the Edwards-Anderson order parameter, ($q_{EA}$)
of the glassy model \cite{binder}. This simple observation will be
useful in the next section, when the results of a lift of the $r$
times degeneracy are discussed.

\section{Density of states and Flory-Huggins series}

The reader may have already noticed that the only effect of the
addition of $r$ times degeneracy to the (ordered or glassy)
Blume-Capel model is the simple relation
\begin{equation}
exp(-\beta D) \to r \  exp(-\beta D)
\end{equation}
i.e.,  one can solve the original model with the \emph{temperature
dependent rescaling} of $D$
\begin{eqnarray}\label{eq:scale}
 D \rightarrow D-T\ln r.
\end{eqnarray}
This is not an incident or an artifact of an approximation
(infinite range model, replica trick) but an exact result. In
fact, for any microscopic configuration of the spin system there
is an excess entropy $\Delta S$ associated  with the $r$ times
degeneracy of any "open" spin, i.e.,
\begin{equation}
\Delta S = r^{\sum_i S_i^2}.
\end{equation}
Correspondingly, the free energy $E-TS$ of an $r$ times degenerate
Blume-Capel model is equivalent to the free energy of the
original, nondegenerate, BC model with the rescaling given by
(\ref{eq:scale}).

A very similar argument has been used by Flory and Huggins in their
discussion of miscibility of polymer melts \cite{flory}
\cite{Huggins}. In the Flory-Huggins theory, if the filling fraction
of one polymer is $\phi$ [and, accordingly,  the fraction of the
other polymer is ($1-\phi$)] the free energy of the system is
written as
\begin{equation} \label{fh}
\beta f = \phi ln(\phi) + (1-\phi)ln(1-\phi) + \chi \phi (1-\phi).
\end{equation}
While the first two terms measure the entropy associated with the
mixture, the last term stands for the energy associated with the
blend. The Flory-Huggins $\chi$ parameter, however, depends on
temperature:
\begin{equation}
\chi = A + \frac{B}{T} + \frac{C}{T^2}
\end{equation}
where the constants A, B and C are  determined experimentally.
Clearly, only the constant B is a "real" interaction parameter, as
it does not depend on the temperature. The other, temperature
dependent constants reflect the "residual entropy" associated with
the interaction between polymers; for example, if a  polymer of one
species tends to take a more compact shape when it is surrounded by
polymers of the opposite species, the corresponding contribution to
the entropy is not included in the first two terms of (\ref{fh}),
but in one of the T-dependent factors A or C.

One may easily identify the shift $D \to D - T ln(r)$ with the A
parameter of the Flory-Huggins series, so this contribution comes
from an \emph{exact} degeneracy of  states. In terms of the
modified Blume-Capel model one recognizes  the other parameter, C,
as related to the finite width of the density of states
distribution maxima.  What happens if the degeneracy of the $r$
"open" states is not exact? In order to consider this problem, let
us assume that there are $r$ interacting states in an interval of
width $\Delta$ centered at  $S= \pm 1$. The case $\Delta = 0 $
corresponds to the exact degeneracy as before, and we are
interested in the corrections to the effective Hamiltonian for a
small interval width.

\begin{figure}
  \includegraphics[width=7.7cm]{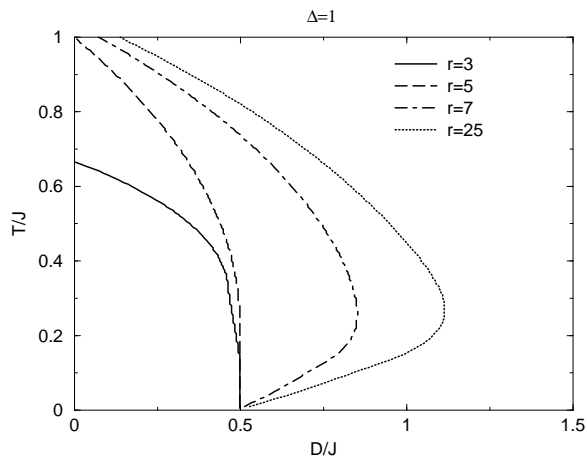}\\
  \caption{Phase diagram for a modified, mean field, version of the ordered
  model in the D-T plane.  There is no exact degeneracy of the "interacting" states. Instead,
  $r$ spin states are equally distributed around the  $\pm 1 $ state,
  with level spacing $\Delta/r$. As seen in the figure, larger
  density of  spin states corresponds to more pronounced inverse
  melting.} \label{resr}
\end{figure}

We begin with numerical examples. In Figure (\ref{resr}) the phase
diagram is shown, for the "almost degenerate" Blume-Capel model,
with $\Delta = 1$ for different values of $r$. As expected, higher
values of $r$ imply more pronounced inverse inverse melting
phenomena and an increase of the ferromagnetic region, since the
"active" spin state is favored by entropy. Of course, all the
lines encounter at the same point for $T=0$, where entropy has no
effect on the state of the system.

\begin{figure}
  \includegraphics[width=7.7cm]{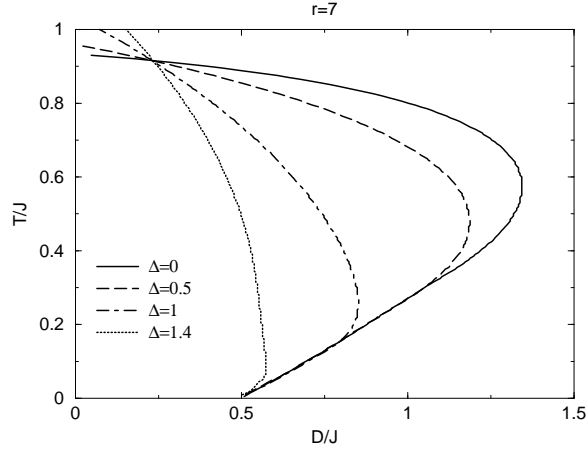}\\
  \caption{The same as Figure (\ref{resr}), but with a constant number of spin levels
  ($r=7$)
  of different widths $\Delta$ around $S=1$. Again larger density of spin (hence energy) states
  implies inverse melting. The occurrence of \emph{two} common points for all curves is discussed in the
  text.}
       \label{resDel}
\end{figure}

In Figure (\ref{resDel}), on the other hand, $r$ is kept constant
while $\Delta$ changes from zero (degenerate BC model) to $1.4$.
The inverse melting manifestation is stronger for the degenerate
case and weakened as $\Delta$ increases. Interestingly, all curves
cross at two points in the $D-T$ plane: one is the point at $T=0$,
where only energetic consideration are important and the system
crosses from the zero spin state to the \emph{maximal} spin state.
As the temperature increases, the phase transition involves finite
populations of other (not maximal) spin states, in order to
increase the entropy of the ferromagnet. At the point where (below
the transition) all active spin states are equally populated there
is no significance of the value of $\Delta$, and all curves
converge again at the same point.

Before making an explicit theoretical considerations, let us make
a distinction between one particle and interaction degeneracies.
If a small perturbation lifts the $r$-fold degeneracy of the spin,
it may come from one of two sources, namely, an intrinsic, "one
particle" splitting (e.g., the polymer in its open conformation
admits many spatial configuration, each of them with slightly
different energy) and an interaction splitting (e.g., the
energetic differences between various conformations of a single
polymer are negligible, but the splitting is induced by the
different energies associated with the relative conformations of
two interacting polymers). In our Blume-Capel Hamiltonian, the
first, single particle situation implies degenerate exchange term
while the second, interacting situation corresponds to degenerate
lattice splitting term.

Let us consider the single particle situation. Here one should
replace any trace on internal single-spin degrees of freedom by
the summation:
\begin{equation}
\sum_{k=-r/2}^{r/2} exp[-\beta D (1+k \frac{\Delta}{r})^2] \approx
\frac{r}{\Delta} \int_{-\Delta/2}^{\Delta/2} e^{-\beta D (1+x)^2} dx
\end{equation}
This integral yields some error function, but we are interested in
the small $\Delta$ corrections to (\ref{eq:scale}). To order
$\Delta^2$, these are:
\begin{equation}\label{eq:scale1}
 D \rightarrow D + \frac{D \Delta^2}{12} -T\ln r  - \frac{ D^2
 \Delta^2}{6T}
\end{equation}
These correction are to be identified with the Flory-Huggins
constants, i.e., as long as the energy associated with the finite
width of the density of states maxima, $\beta D \Delta$, is
smaller then the thermal energy $k_B T$ and one writes the
Flory-Huggins $\chi$ parameter as an infinite series in inverse
powers of $T$, where the actual parametrization of Flory and
Huggins corresponds to the first three terms in this series. As
long as the main contribution to the splitting in the DOS  maxima
comes from a single particle, this argument is applicable to the
ordered system, as well as the disordered one, and to all orders
in the replica symmetry breaking procedure.

The situation changes when the splitting comes from different
exchange interactions associated with various microscopic
conformations of the "open" states. Here one should make a
distinction between the ordered and the disordered states, and a
possibility of deviations from a Flory-Huggins like series.

The simplest case is the ordered one, where now the trace over
single particle states involves the summation:
\begin{equation}
\sum_{k=-r/2}^{r/2} cosh[\beta J m (1+k \frac{\Delta}{r})] \approx
\frac{r}{\Delta} \int_{-\Delta/2}^{\Delta/2} cosh[\beta J (1+x)] dx
\end{equation}
Preforming the integration and expanding the result for small
$\Delta$ one finds, to the leading order in $\Delta$, the
rescaling of D:
\begin{equation}\label{eq:scale2}
 D \rightarrow D  -T\ln r  - \frac{J^2 m^2 \Delta^2}{24 T}
\end{equation}
Notice that the small parameter in the series is the relation
between the energy splitting due to the effective field, $J m
\Delta$, and the energy of  temperature fluctuations $k_B T$. As
long as the system is paramagnetic, i.e., the order parameter $m$
vanishes, there is no effect of this type of splitting at all.
Accordingly, the exchange splitting has no effect on the  location
of second order transition line  and the paramagnetic spinodal
line where the order parameters disappear.

It is important to note the difference, for interaction splitting,
between the ordered and  the disordered case. Looking at the
replica symmetric solution where the trace over single spin
configurations is taken, Eq. (\ref{eq:17} - \ref{eq:10}) one
clearly recognizes another term that contributes to the rescaling
of $D$, even in the paramagnetic phase: this is the term
$exp(\frac{\beta^2 J^2(p-q)}{2}\sum S^2)$ in (\ref{eq:17}) that,
for $q=0$, yields the following corrections:
\begin{equation}\label{eq:scale1}
 D \rightarrow D -T\ln r  + \frac{J^2 \Delta^2 p}{12 T} \left( 1-\frac{2
 J^2 p^2}{T^2} \right).
\end{equation}
This result, again, holds for any order in the replica symmetry
breaking series as long as none of the $q(x)$ parameters differ
from zero. Moreover, even if the glass order parameter takes
finite value, and to any order in the RSB process, the only change
in this expression is the replacement of $p$ by $p-q_{EA}$, as
explained in the previous section.

The intuition beyond this result is simple: in the ordered,
infinite range interaction Blume Capel model there is no local
field from the exchange interaction on a spin as long as the
system is in its paramagnetic state. In the disordered system, on
the other hand, clusters of spins are formed, even above the
transition. While below the transition these clusters are frozen,
above the transition they oscillate coherently at long times, so
the $q$ parameters remains zero, but the spins tend to be in the
interacting state instead of at the zero state. As a result, the
parameter $p$ that measures the tendency of a spin to be in the
interacting state takes finite values even above the transition,
and there is a corresponding local field, $ J \sqrt(p)$, that
"pushes" the spin out of the zero state to either the plus or the
minus state. The new measure for the degeneracy is the ratio
between this local field splitting, $ J \sqrt(p) \Delta$,  and the
temperature smearing. Turning to the polymer analogy, even before
the gelation where the system is not yet frozen, one expects the
polymers to have a tendency for the open conformation. At this
stage, the free energy of the system is effected by tiny
differences in the inter polymer interactions associated with
different spatial conformations, although there is no global
freezing. As $p$ itself is temperature dependent, deviations from
Flory-Huggins behavior are  expected \cite{mendez}  and
non-integer powers appear in the inverse temperature series.

\section{concluding remarks}

There is a difference between the "objective", thermodynamic
definition about \emph{order} and the subjective perception of
this concept. The objective measure for order and disorder is the
entropy of the system. For an unbounded system this quantity
monotonically increases with temperature, giving rise to the
definition of temperature as a measure of the disorder and
fluctuations in the system.

Subjectively, however, one associates \emph{order} with
crystalline structure, frozen molecules or phase separation. These
features may be only part of the global pictures, leading to the
concept of larger "order parameter" as temperature increases. As
reviewed in this paper, this situation does happens in many
physical systems, and then one speaks about inverse melting or
inverse freezing.

We believe that the basic ingredients that appear in the "minimal"
model presented here, namely, a degenerate, entropically favored
interacting state and energetically favored noninteracting state
appear in almost all the physical system that show inverse melting
or inverse freezing. The degenerate Blume-Capel model presented
here, along with its random exchange generalization, supply a
basic framework within which some of the basic qualitative
features of all these systems are demonstrated.

In a generic system, an exact degeneracy of the density of states
never occurs, there is only a peak in the density of states,
corresponding to almost degenerate microscopic states. As shown
here, there is a distinction between a "single particle" (almost)
degeneracy, like the one associated with various conformations of
a polymer, and a "many body" entropically favored states. In the
first case, a Flory-Huggins like theory may be constructed, with
an effective parameter (corresponding to the $\chi$ parameter of
the polymer blends theory) that reflects the effect of the
entropy. In the second case, this Flory-Huggins like description
fails, but the entropy has no effect in the "disordered"
(paramagnetic) phase.

\section{acknowledgements}

The authors wish to thank Jefferson Arenzon, Pablo Debenedetti,
Yizhak Rabin, David Mukamel  and Georg Foltin for most helpful
discussions and comments. This work was supported by the Israel
Science Foundation (ISF) and  by Y. Horowitz Association.

\end{document}